\documentclass[10pt]{article}
\usepackage{amsfonts, amssymb, amsthm,amsmath}

\newcounter{orange}

\newcommand{\be}{\begin{equation}}
\newcommand{\ee}{\end{equation}}
\newcommand{\bea}{\begin{eqnarray}}
\newcommand{\eea}{\end{eqnarray}}

\begin{document}

\begin{center}
{\LARGE{\bf{Time -- of -- arrival operator on the circle.  (Variations on two Mielnik's works)}}}
\end{center}

\begin{center}
\vskip0.25cm

{\bf M. Przanowski \footnote{E-mail address: przan@fis.cinvestav.mx}}

\vskip0.25cm
{\em  Institute of Physics}\\ {\em Technical University of Lodz}\\ {\em  Wolczanska 219, 90-924 Lodz}\\
{\em Poland}

{\bf M. Skulimowski \footnote{E-mail address: mskulim@uni.lodz.pl}} 

\vskip0.25cm
{\em Faculty of Physics and Applied Informatics}\\
 {\em University of Lodz}\\
{\em Pomorska 149/ 153, 90-236 Lodz}\\
{\em Poland}

{\bf J. Tosiek \footnote{E-mail address: tosiek@p.lodz.pl}}

\vskip0.25cm
{\em  Institute of Physics}\\ {\em Technical University of Lodz}\\ {\em  Wolczanska 219, 90-924 Lodz}\\
{\em Poland}

\end{center}



\begin{center}
{\it Dedicated to Bogdan Mielnik}
\end{center}


\begin{abstract}
Using the orthodox Weyl -- Wigner -- Stratonovich -- Cohen  (WWSC) quantization rule we construct a time -- of -- arrival operator for a free particle on the circle. It is shown that this operator is self -- adjoint but the careful analysis of its properties suggests that it cannot represent a `physical' time -- of -- arrival observable. The problem of a 
time -- of -- arrival observable for the `waiting screen' is also considered. A method of avoiding the quantum Zeno effect is proposed and the positive operator valued measure (POV -- measure) or the generalized positive operator valued measure (GPOV -- measure) describing quantum time -- of -- arrival observable for the waiting screen are found. 
\end{abstract}

\section{Introduction}

We begin with some sentences by St. Augustine taken from his `Confessions' \cite{stau}: `What, then, is time? If no one ask of me, I know; if I wish to explain to him who asks, I know not' (Book 11, chapter XIV) and 

`When, therefore, they say that things future are seen, it is not themselves, which as yet are not (that is which are future); but their causes or their signs perhaps are seen, the which already are. Therefore, to those already beholding them, they are not future, but present, from which future things conceived in the mind are foretold.' (Book 11, chapter XVIII).

A shadow of these two phrases can be easily recognized in two contemporary works. One of them, `"Time operator": the challenge persists' by Bogdan Mielnik and Gabino Torres -- Vega \cite{2} shows essential difficulties with a definition of a quantum time observable and the authors conclude that: `While the future of the subject is unknown, it becomes clear, that all intents to obtain the {\it time observable} in the orthodox form of a self -- adjoint operator (in spite of the best stratagems to avoid the Pauli theorem [...]) lead to a blind alley. The resulting operators are typically plagued by some little but persistent difficulties which might look accidental; besides they all suffer some basic defect which seems common for the whole family.' (\cite{2}, p.90)

The main question of the second work `The screen problem' by Bogdan Mielnik \cite{3} can be stated as follows: `One of the crucial statements of quantum mechanics is that the state vector contains complete noncontradictory  information about the system' (\cite{3}, p. 1128) so, Mielnik asks, where is the information about the time coordinate of the event of absorption of a wave packet by the waiting screen (see \cite{3}, Fig. 1). 

The problem of understanding time or, in particular, time -- of -- arrival as a quantum observable, and not as a parameter only, has a long history  and a vast bibliography which starts with distinguished works by W. Pauli \cite{4}, Y. Aharonov and D. Bohm \cite{5}, M. Razavy \cite{6}, G. R. Allcock \cite{7}, E. P. Wigner \cite{8}, J. Kijowski \cite{9}, to mention some of them (see also a nice review of this matter by J. G. Muga and C. R. Leavens \cite{10}).

Although a big effort has been done to solve the problem, we are still far from a satisfactory solution. We have no satisfactory time -- of -- arrival operator as it is very clearly stressed in Mielnik's paper \cite{2} and we have no explicit solution of the waiting screen problem described by Mielnik \cite{3}.

The aim of the present work is to study these two questions once more. In Sec. \ref{s2}, using the `orthodox' Weyl -- Wigner -- Stratonovich -- Cohen (WWSC) quantization rule we find a time -- of -- arrival operator for a free particle on a circle. It is shown that this operator has nice mathematical properties, namely it is bounded, self -- adjoint and of Hilbert -- Schmidt type. However, it cannot be interpreted as the operator representing the physical time -- of -- arrival observable since it is `plagued by some little but persistent difficulties.' In Sec. \ref{s3} we consider the waiting screen (detector) problem for a free particle. Using the `orthodox' reduction of state assumption in quantum mechanics and avoiding the quantum Zeno effect we find a formula for the average time -- of -- arrival, which in turn defines the {\it (generalized)  positive operator valued measure} $\big($(G)POV -- measure$\big)$.

Our considerations are similar to the ones related to the decoherent histories approach to quantum mechanics developed by J.J. Halliwell and J. M. Yearsley \cite{11}, \cite{12}.

The present paper has, in fact, the form of two variations on the themes given by Mielnik in his works \cite{2} and \cite{3} and it is an honour and a great pleasure to dedicate these variations to Bogdan Mielnik on the occasion of his 75th birthday.

\section{Time -- of -- arrival operator on the circle}
\label{s2}
Consider first a free particle on the $x$ -- axis. If the coordinate of the particle at  the initial moment $t_0=0$ is $x$ then the time of arrival of this particle at the point $X=0$ (screen) in classical mechanics reads
\be
\label{1}
T= -  m \frac{x}{p},
\ee
where $m$ is the mass of the particle and $p$ stands for its momentum. Quantization of (\ref{1}) in the symmetric ordering leads to the {\it Aharonov  -- Bohm time -- of -- arrival operator} \cite{5}
\be
\label{2}
\hat{T}= - \frac{m}{2} \left(\hat{x} \hat{p}^{-1}+  \hat{p}^{-1} \hat{x}\right)= - i \frac{m}{\hbar} \frac{1}{\sqrt{k}}
\frac{d}{dk}\left( \frac{1}{\sqrt{k}} \right), \;\; k = \frac{p}{\hbar}
\ee
which is maximally symmetric but has no self -- adjoint extensions. The natural way out from that difficulty has been found by N. Grot, C. Rovelli and R. S. Tate \cite{13}, and it consists in an appropriate regularization of the operator (\ref{2}) in a small neighbourhood  of the singular point $k=0.$ Thus one gets the {\it regulated time -- of -- arrival operator} 
\be
\label{3}
\hat{T}_{\varepsilon }= - i \frac{m}{\hbar} \sqrt{f_{\varepsilon }(k)} \frac{d}{dk}\left( \sqrt{f_{\varepsilon }(k)} \right),
\ee
where $\varepsilon >0$ is an arbitrary small positive number and $f_{\varepsilon }(k)$ is a real bounded continuous function such that
\be
\label{4}
f_{\varepsilon }(-k) = - f_{\varepsilon }(k)\;\; , \;\; f_{\varepsilon }(k)= \frac{1}{k} \; {\rm for} \; |k|> \varepsilon \;, \; \forall_{ k \neq 0 }\; f_{\varepsilon }(k) \neq 0 
\ee
(for instance $f_{\varepsilon }(k)= \frac{1}{k} \; {\rm for}\; |k|> \varepsilon $ and 
$ f_{\varepsilon }(k)= \frac{k}{\varepsilon^2 } \; {\rm for} \;|k|< \varepsilon$). It has been shown in \cite{13} that
$\hat{T}_{\varepsilon }$ is self -- adjoint. This is a very good news. However, there are also bad news:
\begin{enumerate}
\item
J. Oppenheim, B. Reznik and W. G. Unruh \cite{14} have shown that if the particle is in an eigenstate of $\hat{T}_{\varepsilon }$ corresponding to some eigenvalue $\tau $ of $\hat{T}_{\varepsilon },$ then at the moment $\tau$ i.e. at the predict time of arrival at the screen this particle can be detected far away from the screen with probability $\frac{1}{2}$
\item
\label{i2}
Eigenstates $| \tau, \pm > $ (note the degeneration !) of $\hat{T}_{\varepsilon }$ are not {\it time -- translation invariant} i.e.
\be
\label{5}
\exp \left\{ - \frac{i}{\hbar} \frac{\hat{p}^2}{2m} t \right\} | \tau, \pm > \neq | \tau-t , \pm >
\ee
(This is a consequence of Pauli's theorem \cite{4})
\item
\label{i3}
Eigenvalues $\tau$ of $\hat{T}_{\varepsilon }$ can be both positive and negative. It seems that from the experimental point of view the negative time of arrival, $\tau < 0,$ is questionable in quantum mechanics.
\end{enumerate}
The mentioned above points show that one can hardly consider $\hat{T}_{\varepsilon }$ as a correct time -- of -- arrival operator. Our first idea is to avoid the objection \ref{i3}. 

To this end we propose to deal with a free particle on the circle. Let $- \pi < \Theta \leq \pi$ denote the angle coordinate of a particle at the moment $t=0$ on the circle of radius $r$ and let $L$ be the angular momentum of the particle. Then the time of arrival of this particle at the point $\Theta=0$ (screen) is given by the following function
\be
\label{6}
T(\Theta,L)= m r^2 \cdot \left\{ 
\begin{array}{ccl}
- \frac{2 \pi + \Theta}{L} & {\rm for} & \Theta < 0\; ,\; L<0  \vspace{0.2cm} \\ \vspace{0.2cm}
- \frac{ \Theta}{L} & {\rm for} & \Theta > 0\; ,\; L<0 \; {\rm or} \; \Theta <0\; ,\; L>0 \\ \vspace{0.2cm}
\frac{2 \pi - \Theta}{L} & {\rm for} & \Theta > 0\; ,\; L>0 \\
g(\Theta) \geq 0 & {\rm for} & L=0
\end{array}
\right.
\ee
Of course $T(\Theta,L)$ describes the {\it first -- passage time} \cite{10}.

An arbitrary nonnegative function $g(\Theta) \geq 0$ plays the analogous role as the function $f_{\varepsilon }(k)$ in (\ref{3}), i.e. $g(\Theta)$ regularizes the classical function $T(\Theta,L)$ at the point $L=0.$ We quantize $T(\Theta,L)$ according to the WWSC method \cite{15} -- \cite{18}. Thus we arrive at the operator 
\be
\label{7}
\hat{T}_{({\bf K} )}= 
\sum_{n= - \infty}^{\infty} \int_{- \pi}^{\pi} T(\Theta, n \hbar) \:\hat{\Omega}_{({\bf K})}(\Theta, n ) \:\frac{d \Theta}{2 \pi}
\ee
where $\hat{\Omega}_{({\bf K})}(\Theta, n )$ 
is the {\it generalized Stratonovich -- Weyl quantizer, } which in the case of a circle reads \cite{19} -- \cite{23}
\be
\label{8}
\hat{\Omega}_{({\bf K})}(\Theta, n )= \sum_{l= - \infty}^{\infty} \int_{- \pi}^{\pi} {\bf K} (\sigma, l)
\exp \left\{ -i(\sigma n + l \Theta)\right\} \hat{U}(\sigma, l) \frac{d \sigma}{2 \pi}
\ee
with
\[
\hat{U}(\sigma, l)= \exp \left\{ - \frac{i}{2} l \sigma \right\} \exp \left\{  \frac{i}{\hbar}  \sigma \hat{L}\right\}
\exp \left\{ i l \hat{\Theta} \right\}=
\]
\be
\label{9}
\exp \left\{  \frac{i}{2} l \sigma \right\} 
\exp \left\{ i l \hat{\Theta} \right\}
\exp \left\{  \frac{i}{\hbar}  \sigma \hat{L}\right\}
= \sum_{k= - \infty}^{\infty}\exp \left\{ i\left( k+ \frac{l}{2}\right) \sigma \right\} |k+l > < k|
\ee 
where $|k>, k=0, \pm 1, \ldots,$ stands for the normalized eigenvector of $\hat{L}$
\be
\label{10}
\hat{L}|k>= k \hbar |k>\; , \; <k|k'>= \delta_{k k'}.
\ee
The kernel function ${\bf K} = {\bf K} (\sigma, l), \; - \pi < \sigma \leq \pi, \; l \in {\mathbb Z}$ determines an ordering of operators. For example if ${\bf K}=1$ then one gets the {\it Weyl ordering}, for ${\bf K}= \cos \left( \frac{l \sigma}{2}\right)$ one obtains the {\it symmetric ordering}. Therefore, using (\ref{7}), (\ref{8}) and 
(\ref{9}) with ${\bf K}= \cos \left( \frac{l \sigma}{2}\right)$ and performing simple but rather tedious manipulations we find the time -- of -- arrival operator in the symmetric ordering for a free particle on the circle
\[
\hat{T}_S= m r^2 \cdot \Bigg\{\frac{1}{2i \hbar} \sum_{\tiny \begin{array}{c}j,k=- \infty,\\ j \neq 0,\:k \neq 0, \\ j \neq k 
\end{array}}^{\infty} \frac{j+k}{jk (j-k)} |j><k| + \frac{\pi}{\hbar}\sum_{\tiny \begin{array}{c} k=-\infty \\ k \neq 0 \end{array}}^{\infty} \frac{1}{|k|} |k><k|+
\]
\[
+ \sum_{\tiny \begin{array}{c}k=- \infty,\\ k \neq 0
\end{array}}^{\infty} \left[ \frac{1}{2i \hbar k^2} + \frac{1}{4 \pi} \int_{- \pi}^{\pi} g(\Theta) \exp \{-ik \Theta \}d \Theta
\right] |k><0|+
\]
\[
+ \sum_{\tiny \begin{array}{c}k=- \infty,\\ k \neq 0
\end{array}}^{\infty} \left[- \frac{1}{2i \hbar k^2} + \frac{1}{4 \pi} \int_{- \pi}^{\pi} g(\Theta) \exp \{ik \Theta \}d \Theta
\right] |0><k|+
\]
\be
\label{11}
\left. +  \frac{1}{2 \pi} \int_{- \pi}^{\pi} g(\Theta) d \Theta |0><0| \right\} .
\ee
One can show that the operator $\hat{T}_S$ has nice mathematical properties. It is defined on the all Hilbert space $L^2(S^1).$ Then it is self -- adjoint, bounded and of Hilbert -- Schmidt type so it is also a completely continuous (compact) operator. Hence, due to the Hilbert -- Schmidt theorem  $\hat{T}_S$ can be represented as follows
\be
\label{12}
\hat{T}_S= \sum_{k=1}^{\infty} \tau_k |\tau_k><\tau_k|,
\ee
\[
\tau_k \in {\mathbb R}\; ,\;  \sum_{k=1}^{\infty} \tau_k^2 < \infty\;, \; <\tau_k|\tau_l>= \delta_{kl}\;,\; 
\sum_{k=1}^{\infty} |\tau_k><\tau_k|= \hat{1}.
\]
One can also show that the {\it time -- of -- arrival operator in the Weyl ordering} has the same properties. We expect that these properties will be recovered for any time -- of -- arrival operator of a free particle on the circle which is constructed by quantizing some classical time -- of -- arrival function corresponding to the {\it  first -- passage time}.

Further analysis of the properties of the time -- of -- arrival operator $\hat{T}_S$ leads to the conclusions
\renewcommand{\theenumi}{(\alph{enumi})}
\begin{enumerate}
\item
\label{aa1}
$\hat{T}_S$ has a discrete spectrum with the accumulation point $0.$ For every $\lambda>0$ there exists a finite number of eigenvalues $\tau_k$ of $\hat{T}_S$ such that $|\tau_k|> \lambda.$ The spectrum of $\hat{T}_S$ depends on the mass of the particle, what means, for instance, that the participants of the Bialowieza conference are not able to arrive at Bialowieza at the same time. Moreover, we should consider the `clock time' which appears to be continuous and the arrival time which for a given particle is discreet.
\item
In general
\be
\label{13}
\exp \left\{ -\frac{i }{\hbar} t \hat{H}\right\}|\tau_k> \neq {\rm const.}\: \cdot |\tau_k - t>
\ee
for any Hamiltonian $\hat{H}$ i.e. $\hat{T}_S$ is not a {\it time -- translation invariant} (compare with (\ref{5})).
\item
\label{aa3}
Numerical (computer) results show that for $g(\Theta)= {\rm const.}\geq 0$ even so the classical  time -- of -- arrival function $T(\Theta,p) \geq 0$ the operator $\hat{T}_S$ has positive as well as negative eigenvalues. The remedy for this could be the definition of the time -- of -- arrival operator $\sqrt{\hat{T}_S}$ (see also \cite{11}, \cite{12}). However, this does not cure the lack of the time -- translation invariance.
Moreover, our preliminary calculations lead to the arguments analogous to those by J. Oppenheim, B. Reznik and W. G. Unruh \cite{14} i.e. assuming $g(\Theta)= {\rm const.} \geq 0,$ at the predict time of arrival the particle can be detected far away from the point $\Theta=0$ (screen) with considerable probability.
\end{enumerate}
 Most likely the statements \ref{aa1} -- \ref{aa3} hold true for any time -- of -- arrival operator constructed by the WWSC method from a classical time -- of -- arrival function for a free particle on the circle. Although further investigations for an arbitrary $g(\Theta)$ are needed (we are working on this problem) one can repeat Mielnik's and Torres -- Vega's words: ` "Time operator" The challenge persists'.

Moreover, in contrary to some suggestions \cite{13}, \cite{24} it seems that one cannot `forget time' and that $x$ and $t$ cannot be treated on equal footing in quantum mechanics.
\section{A waiting screen}
\label{s3}
Here we deal with a particle in ${\mathbb R}^3$ which can be detected by a {\it waiting screen (detector)}. We assume that the particle is absorbed (detected) if and only if it falls into some domain $V \subset {\mathbb R}^3.$ Define two projectors
\be
\label{14}
\hat{E}:= \int_{V} |\vec{x}>dx^3<\vec{x}|\; ,\; \hat{E}'= \hat{1}- \hat{E}.
\ee
Consider then a time interval $[0,t]$ and choose the moments of time $0=t_0 < t_1< \ldots <t_n=t.$ If the initial state of a particle is $|\Phi_{in} >\;, \; <\Phi_{in} |\Phi_{in} >=1, $ and one assumes the orthodox doctrine of quantum mechanics about a state reduction also for measurements performed  without touching the object, then straightforward calculations show that the probability ${\bf P}_j\: ,\: j=0,1,\ldots, n,$ that the particle will be absorbed at the moment $t_j$ reads
\[
{\bf P}_j= <\Phi_{in}| \hat{E}' \exp \left\{ \frac{i }{\hbar} (t_1- t_0) \hat{H}\right\}
\hat{E}' \exp \left\{ \frac{i }{\hbar} (t_2- t_1) \hat{H}\right\} \ldots
\]
\[
\hat{E}' \exp \left\{ \frac{i }{\hbar} (t_j- t_{j-1}) \hat{H}\right\}
\hat{E} \exp \left\{ - \frac{i }{\hbar} (t_j- t_{j-1}) \hat{H}\right\}\hat{E}' \ldots
\]
\be
\label{15}
 \exp \left\{ -\frac{i }{\hbar} (t_2- t_1) \hat{H}\right\} \hat{E}'
\exp \left\{- \frac{i }{\hbar} (t_1- t_0) \hat{H}\right\} \hat{E}'
|\Phi_{in}>
\ee
where $\hat{H}$ is the Hamiltonian (see B. Misra and E. C. G. Sudarshan \cite{25} and \cite{11}, \cite{12}).

Taking $t_j - t_{j-1}= \frac{t}{n}, \; j=1, \ldots, n$ we obtain 
\be
\label{16}
{\bf P}_j= <\Phi_{in}|\left(  \hat{E}' \exp \left\{ \frac{i }{\hbar} \frac{t}{n} \hat{H}\right\}\right)^j
\hat{E} \left(
\exp \left\{- \frac{i }{\hbar} \frac{t}{n}  \hat{H}\right\} \hat{E}'\right)^j
|\Phi_{in}>.
\ee
If $\hat{H}$ is self -- adjoint and nonnegative then \cite{25}, \cite{26}
\be
\label{17}
\lim_{n \rightarrow \infty}{\bf P}_n=0.
\ee
This is, of course, the famous {\it quantum Zeno effect} which in our case states that if the particle is not absorbed at the moment $t_0=0$ then it will not be absorbed at all. To avoid this paradoxical statement one can assume that
\renewcommand{\theenumi}{(\Roman{enumi})}
\begin{enumerate}
\item
$\hat{E}'$ is not a projector $\hat{1}- \hat{E}.$ This corresponds to the assumption that there exists a complex potential \cite{7}, \cite{11}, $-iV_0, \; V_0 > 0,$ such that 
\be
\label{18}
\hat{E}'= \exp \left\{- \frac{i }{\hbar} \frac{t}{n}  (-i V_0 \hat{E})\right\}= \hat{1} - \hat{E} + 
\exp \left\{- \frac{V_0 }{\hbar} \frac{t}{n} \right\}\hat{E}\;\; ,\;\; \frac{V_0 }{\hbar} \frac{t}{n} \gg 1.
\ee
\end{enumerate}
Alternatively one can assume that only a {\it  partial state reduction } has place when the measurement without interaction is performed and $\hat{E}'$ describes such a partial state reduction $(\hat{E}' \cdot \hat{E}' \neq \hat{E}').$
\begin{enumerate}
\addtocounter{enumi}{1}
\item
Continuous measurement is not allowed and
\be
\label{19}
\eta:= \frac{t}{n} > \tau_z= \left(<\Phi_{in}| \hat{H}^2|\Phi_{in}> - (<\Phi_{in}| \hat{H}|\Phi_{in}>)^2 \right)^{-\frac{1}{2}} \hbar
\ee
where $\tau_z$ is the {\it Zeno time}. 
\end{enumerate}
Note that B. Mielnik considers in \cite{3} this last argument as `visibly unfair' (see \cite{3} p. 1123). We guess that it is not so unfair if one takes into account that any measurement device has a specific dead time. The assumption (\ref{19}) is given also in \cite{11}, \cite{12}.

Suppose that
\be
\label{20}
\sum_{j=0}^{\infty} {\bf P}_j= 1
\ee
where from (\ref{15}) with (\ref{19}) we have
\be
\label{21}
{\bf P}_j= <\Phi_{in}|\left(  \hat{E}' \exp \left\{ \frac{i }{\hbar} \eta \hat{H}\right\}\right)^j
\hat{E} \left(
\exp \left\{- \frac{i }{\hbar} \eta \hat{H}\right\} \hat{E}'\right)^j
|\Phi_{in}>.
\ee
Then the {\it average time -- of -- arrival} reads
\be
\label{22}
<\tau>= \sum_{j=0}^{\infty}j \eta {\bf P}_j.
\ee
Therefore, one arrives at the conclusion that in the present case quantum time -- of -- arrival is defined by the {\it positive operator valued measure} (POV -- measure)
\be
\label{23}
{\mathbb N} \ni j \longmapsto  \left(  \hat{E}' \exp \left\{ \frac{i }{\hbar} \eta \hat{H}\right\}\right)^j
\hat{E} \left(
\exp \left\{- \frac{i }{\hbar} \eta \hat{H}\right\} \hat{E}'\right)^j \;\;  {\mathbb N}= \{0,1,\ldots\}
\ee
(About POV -- measures see e.g. \cite{27}, \cite{28}). If (\ref{20}) does not hold i.e. the particle can be not absorbed at all then $\sum_{j=0}^{\infty} {\bf P}_j < 1$ and the average time of arrival can be defined as
\be
\label{24}
<\tau>= \frac{\sum_{j=0}^{\infty}j \eta {\bf P}_j}{\sum_{j=0}^{\infty} {\bf P}_j}
\ee
and, consequently, the formula (\ref{23}) gives now a {\it generalized positive operator valued measure} (GPOV -- measure).

All results given hitherto in this section can be easily generalized on the case of a particle moving on some submanifold of ${\mathbb R}^3.$ In particular one can quickly carry over the last result to the case of a particle on the circle with the waiting screen. Assuming now that we deal with multiple crossing the screen by the particle we can state that (\ref{20}) holds true and, consequently, quantum time -- of -- arrival for a particle on the circle is given by the POV -- measure (\ref{23}).

Finally, according to our considerations, a partial answer to the question asked by Mielnik in \cite{3} could be the following: {\it The information about the time coordinate of the event of absorption of a wave packet by the waiting screen is contained in the formula (\ref{22}) or, in general, in (\ref{24})}.

\section*{Acknowledgements}
M. P. and J. T. were partially supported by the CONACYT (Mexico) grant No 103478.


\begin{thebibliography}{11111}
\bibitem{stau}
 St. Augustine, \\ http://www.leaderu.com/cyber/books/augconfessions/bk11.html 
 
 \bibitem{2}
 B. Mielnik and G. Torres -- Vega, {\it "Time operator": The challenge persists}, Concepts of Physics {\bf II} (2005), 81 -- 97.
 
 \bibitem{3}
 B. Mielnik, {\it The screen problem}, Found. Phys. {\bf 24} (1994), 1113 -- 1129.
 
 \bibitem{4}
 W. Pauli, in S. Fl\"{u}gge (Ed.), {\it Encyclopedia of Physics} , vol. {\bf 5} p. 60, Springer, Berlin, Heidelberg, New York, 1958.
 
 \bibitem{5}
 Y. Aharonov and D. Bohm, Phys. Rev. {\bf 122} (1961), 1649.
 
 \bibitem{6}
 M. Razavy, Nuovo Cim. {\bf 63B}, (1969), 271.
 
 \bibitem{7}
 G. R. Allcock, Ann. Phys. {\bf 53} (1969) 253; Ann. Phys. {\bf 53} (1969) 286; Ann. Phys. {\bf 53} (1969) 311.
 
 \bibitem{8}
 E. P. Wigner, in {\it Aspects of Quantum Theory}, Eds. A. Salam and E. P. Wigner, p. 237,   Cambridge, London, 1972.

\bibitem{9}
J. Kijowski, Rep. Math. Phys. {\bf 6} (1974), 361.

\bibitem{10}
J. G. Muga and C. R. Leavens, Phys. Rep. {\bf 338} (2000), 353.

\bibitem{11}
J. J. Halliwell and J. M. Yearsley, Phys. Rev. A {\bf 79} (2009), 062101. 

\bibitem{12}
J. J. Halliwell and J. M. Yearsley, Phys. Lett. A {\bf 374} (2009), 154. 

\bibitem{13}
N. Grot, C. Rovelli and R. S. Tate, Phys. Rev. A {\bf 54} (1996), 4676.

\bibitem{14}
J. Oppenheim, B. Reznik and W. G. Unruh, Phys. Rev. A {\bf 59} (1999), 1804.

\bibitem{15}
H. Weyl, {\it The Theory of Groups and Quantum Mechanics}, Dover Publications, New York, 1931.

\bibitem{16}
E. P. Wigner, Phys. Rev.  {\bf 40} (1932), 749.

\bibitem{17}
R. L. Stratonovich, Sov. Phys. JETP {\bf 31} (1956), 1012.

\bibitem{18}
L. Cohen, J. Math. Phys. {\bf 7} (1966), 781.

\bibitem{19}
N. Mukunda, Am. J. Phys. {\bf 47} (1979), 182.

\bibitem{20}
M. V. Berry, Phil. Trans. R. Soc. London A {\bf 287} (1977), 237.

\bibitem{21}
P. Kasperkovitz and M. Peev, Ann. Phys. {\bf 230} (1994), 21.

\bibitem{22}
J. F. Pleba\'{n}ski, M. Przanowski and J. Tosiek, Acta Phys. Pol. B {\bf 27} (1996), 1961.

\bibitem{23}
J. F. Pleba\'{n}ski, M. Przanowski, J. Tosiek and F. Turrubiates, Acta Phys. Pol. B {\bf 31} (2000), 561.

\bibitem{24}
C. Rovelli, {\it Forget time}, Essay written for the FQXi contest on the Nature of Time (2008), arXiv:0903.3832 [gr-qc].

\bibitem{25}
B. Misra and E. C. G. Sudarshan, J. Math. Phys. {\bf 18} (1977), 756.

\bibitem{26}
P. Exner, T. Ichinose, H. Neidhardt and V. A. Zagrebnov, Int. Eq. Operator Th. {\bf 57} (2007), 67.

\bibitem{27}
P. Bush, M. Grabowski and P. J. Lahti, {\it Operational Quantum Mechanics}, Springer, Berlin, 1995.

\bibitem{28}
M. Skulimowski, Phys. Lett. A {\bf 297} (2002), 129; Phys. Lett. A {\bf 301} (2002), 361.

\end{thebibliography}
\end{document}